\documentclass[structuredabstrac]{aa}  
\usepackage{bbding}
\usepackage{graphicx}
\usepackage{natbib}
\usepackage{xspace}
\xspaceaddexceptions{]\}}

\usepackage[colorlinks, linkcolor=blue, citecolor=blue, breaklinks=true]{hyperref}  
\usepackage[update,prepend]{epstopdf}
\usepackage{booktabs}
\usepackage{amsmath}

\newcommand{\GILDAS}{\texttt{GILDAS}\xspace}
\newcommand{\CLASS}{\texttt{CLASS}\xspace}

\newcommand{\herschel}{\textit{Herschel}\xspace}

\newcommand{\Cp}{\ion{C}{ii}\xspace}

\newcommand{\HII}{\ion{H}{ii}\xspace}

\newcommand{\NeII}{\ion{Ne}{ii}\xspace}

\newcommand{\HH}    {\mbox{H$_2$}\xspace}           
\newcommand{\thCO}  {\mbox{$^{13}$CO}\xspace}       
\newcommand{\CeiO}  {\mbox{C$^{18}$O}\xspace}       
\newcommand{\CHp}  {\mbox{CH$^+$}\xspace}              
\newcommand{\HHOp}  {\mbox{H$_2$O$^+$}\xspace}      
\newcommand{\OHp}  {\mbox{OH$^+$}\xspace}           

\newcommand{\emm}[1]{\ensuremath{#1}}   
\newcommand{\emr}[1]{\emm{\mathrm{#1}}} 
\newcommand{\unit}[1]{\emm{\, \emr{#1}}}
\newcommand{\K}   {\unit{K}\xspace}

\newcommand{\cmq}{\unit{cm^{-2}}\xspace}
\newcommand{\cmc}{\unit{cm^{-3}}\xspace}

\newcommand{\kms}   {\unit{km\,s^{-1}}\xspace}
\newcommand{\Kkms}{\unit{K\,km\,s^{-1}}\xspace}

\newcommand{\GHz} {\unit{GHz}\xspace}

\newcommand{\pc}    {\unit{pc}\xspace}

\newcommand{\hab}{\emm{G_{\rm 0}}\xspace}
\newcommand{\Tmb}{\emm{T_\emr{mb}}\xspace}
\newcommand{\Tex}{\emm{T_\emr{ex}}\xspace}

\newcommand{\Tkin}{\emm{T_\emr{kin}}\xspace}

\newcommand{\nhh}{\emm{n_\emr{H_2}}\xspace}

\newcommand{\ex}[2]{\ensuremath{#1 \times 10^{#2}}\xspace}

\usepackage{txfonts}

\usepackage{soul}

\begin{document}

\title{Kinematics of the ionized-to-neutral interfaces in Monoceros R2\thanks{Herschel is an ESA space observatory with science instruments provided by European-led Principal Investigator consortia and with important participation from NASA.} }

\author{
P.~Pilleri\inst{1,2,3},
A.~Fuente\inst{2},
M.~Gerin\inst{4},
J.~Cernicharo\inst{3},
J.~R.~Goicoechea\inst{3},
V.~Ossenkopf\inst{5},
C.~Joblin\inst{6,7},
M.~Gonz\'alez-Garc\'ia\inst{8},
S.~P.~Trevi\~no-Morales\inst{8},
\'A.\ S\'anchez-Monge\inst{5,9},
J.~Pety\inst{10},
O.~Bern\'e\inst{6,7},
C.~Kramer\inst{8}
}

 \institute{%
Los Alamos National Laboratory, P.O. Box 1663, Los Alamos (NM) 87545, USA
\and
Observatorio Astron\'omico Nacional, Apdo. 112, E-28803 Alcal\'a de Henares (Madrid), Spain
\and
Centro de Astrobiolog\'ia, (INTA-CSIC), Ctra. M-108, km. 4, E-28850 Torrej\'on de Ardoz, Spain
\and
LERMA, Observatoire de Paris, 61 Av. de l'Observatoire, 75014 Paris, France 
\and
 I. Physikalisches Institut der Universit\"at 
 zu K\"oln, Z\"ulpicher Stra\ss{}e 77, 50937 K\"oln, Germany
\and
 Universit\'e de Toulouse, UPS, IRAP, 9 avenue du colonel Roche, 31062 Toulouse cedex 4, France
 \and
 CNRS, UMR 5187, 31028 Toulouse, France
\and
Instituto de Radio Astronom\'ia Milim\'etrica (IRAM), Avenida Divina Pastora 7, Local 20, 18012 Granada, Spain
 \and
Osservatorio Astrofisico di Arcetri, INAF, Largo E. Fermi 5, I-50125 Firenze, Italy
\and
Institut de Radioastronomie Millim\'etrique, 300 Rue de la Piscine, 38406 Saint Martin d'H\'eres, France
}

\authorrunning {P.~Pilleri, et al.} 
\titlerunning{Kinematics of the ionized-to-neutral interfaces in Monoceros R2}


\abstract
{Monoceros R2 (Mon~R2), at a distance of 830 pc, is the only ultra-compact \HII\ region (UC \HII)  where its associated photon-dominated region (PDR) can be resolved with  the \textit{Herschel} Space Observatory.}
{Our aim is to investigate observationally the kinematical patterns in the interface regions ({\it i.e.,}\ the transition from atomic to molecular gas) associated with Mon~R2.
}
{We used the HIFI instrument onboard \herschel to observe the line profiles of the reactive ions \CHp , \OHp and \HHOp toward different positions in Mon~R2. We derive the column density of these molecules and compare them with gas-phase chemistry models.}
{
The reactive ion \CHp is detected both in emission (at central and red-shifted velocities) and in absorption (at blue-shifted velocities).  
\OHp is detected in absorption at  both  blue- and red-shifted velocities, with similar column densities. \HHOp is not detected at any of the positions, down to a rms of 40 mK toward the molecular peak. At this position, we find 
 that the \OHp absorption originates in a mainly atomic medium, and therefore is associated with the most exposed layers of the PDR. These results are consistent with the 
predictions from photo-chemical models. The line profiles are consistent with the atomic gas being entrained in the ionized gas flow along the walls of the cavity of the \HII region. Based on  this evidence, we are able to propose a new geometrical model for this region. 
}
{The kinematical patterns of the \OHp and \CHp absorption indicate the existence of a layer of mainly atomic gas for which we have derived, for the first time, some physical parameters and its dynamics.}

\maketitle


\section{Introduction}

Monoceros R2 (Mon~R2) is a relatively nearby (830\pc) ultra-compact \HII region (UC\HII). Due to its brightness and proximity, it is the best source  in which to investigate the chemistry and physics of this stage of high-mass stellar formation.
The main ionizing source, IRS1, is located at the center of a spherical cavity free of molecular gas which extends for about 20\arcsec\
(0.08\pc) in radius  \citep{choi00}. The UC\HII is surrounded by several  photon-dominated regions (PDRs)  that are characterized by
different physical and chemical conditions \citep{rizzo03, berne09a, pilleri13}. These PDRs  have a roughly   circular
spatial distribution, with a projected thickness between 4\arcsec\, and 6\arcsec\, \citep[$\sim0.02$\pc,][]{berne09a}.  \citet{rizzo03} estimated the intensity of the UV radiation field  to be $\hab \sim \ex{5}{5}$ in units of the \citet{habing68} field, by fitting the observed far-IR intensity with a blackbody curve, assuming an effective temperature of  $T_{\rm eff} =25.000$\K and using the projected radius of the \HII region as the real distance from the star. \citet{berne09a} studied the \HH mid-IR lines from  {\it Spitzer}-IRS observations and  estimated a  hydrogen density in the PDRs ranging from $n_{\rm H} =\ex{4}{3}$ to \ex{4}{5}\cmc. Higher densities, up to $n_{\rm H} = \ex{6}{6}$\cmc, were found by \citet{choi00}, \citet{rizzo03} and \citet{pilleri13} to account for the emission of molecular species in the far-IR and radio domains. The UC\HII and the PDRs are surrounded by a  moderate density ($\nhh\sim \ex{5}{4}\cmc$), cometary-shaped molecular cloud \citep{fuente10, pilleri12a, pilleri13}. 

Observations and modeling of this source showed that the molecular cloud is relatively quiescent, with a  systemic velocity of the main cloud being $\sim11\kms$ and expansion velocities $\lesssim 1$\kms \citep{fuente10, pilleri12a}. High velocity wings, observed at red-shifted velocities  relative to the cloud systemic velocity, can be attributed either to the expansion of the PDR surrounding the \HII region or to the relic of a now inactive outflow \citep{giannakopoulou97, pilleri12a, pilleri13}. \citet{jaffe03}, through the analysis of high spatial and spectral resolution  observations of the 12.8\,$\mu$m \NeII line, suggested that the innermost \HII region is expanding at a high velocity ${\rm v}_{\rm exp} = 10\kms$. \citet{zhu05} refined this scenario, showing that these patterns are more likely explained by a model in which the cavity is devoid of gas and dust and is maintained by stellar wind pressure. In this interpretation, the central \HII region is not expanding, but  the ionized gas runs along the walls of the surrounding cloud, a behavior also registered in other UC\HII regions \citep{zhu08}. Finally, the overall Mon~R2 complex seems to undergo a slow rotation \citep{loren77, pilleri12a}. 

 Our previous observations of the molecular gas in Mon~R2 \citep{pilleri12a} put a strong upper limit to the expansion velocity of the innermost layers of this region, in agreement with the interpretation of \citet{zhu05}. We have obtained  new  high spectral resolution observations with \herschel that  provide new insights into the geometry and kinematics of the intermediate atomic layers between the \HII gas and the fully molecular gas.
  \herschel allows us to detect reactive ions such as \OHp, \HHOp and \CHp that trace the first steps of the chemistry in UV-illuminated gas.
In particular, the Heterodyne Instrument for the Far Infrared   \citep[HIFI, ][]{degraauw10}  allows us to resolve the line profiles and  constrain the kinematics of the gas where the atomic/ionized to molecular/neutral gas  transition takes place. 
 In Sect. \ref{sec_obs} we present our observations, and the results are reported in Sect. \ref{sec_model}.  Sections \ref{sec_disc}
 and \ref{sec_conclus} present the discussion and conclusions.

\begin{table*} 
\caption{Observed noise and continuum brightness, in \Tmb.} 
\label{table_summary} 
\begin{center} 
\begin{tabular}{l|cc|cc|cc}
\toprule	
			& \multicolumn{2}{c}{IF [0\arcsec, 0\arcsec]}	& \multicolumn{2}{c}{MP1 [10\arcsec, -10\arcsec]}	&\multicolumn{2}{c}{CP1 [-28\arcsec, 10\arcsec]}  \\ 
\midrule
Species		& $rms$	& $T_{\rm c}$	& $rms$	& $T_{\rm c}$						&  $rms$ & $T_{\rm c}$					\\					
			& [mK]	& [K]							& [mK]	& [K]								& [mK]	& [K]			\\
\midrule		
\CHp			& 	25	&	0.45						& 129	& 0.57\tablefootmark{(a)}				& 	 153 &  0.26\tablefootmark{(a)}	 \\
\OHp			& 21		&	0.57						& 60		& 0.72							& 19	&	0.33	\\
\HHOp		& 180	&	 0.67\tablefootmark{(a)}	 & 40		& 0.85				& 280	& 	0.39\tablefootmark{(a)}	 \\	
\bottomrule
\end{tabular} 
\end{center}
\tablefoot{Noise $rms$ values have been calculated on a \Tmb scale with $\Delta v = 0.5\kms$.\\
\tablefoottext{a}{The observed continuum level is not reliable, and the value reported here is an estimate based on the \OHp continuum (see text).}\\}
\end{table*}

\section{Observations}
\label{sec_obs}

\begin{figure}
\centering
\includegraphics[width=0.9\linewidth]{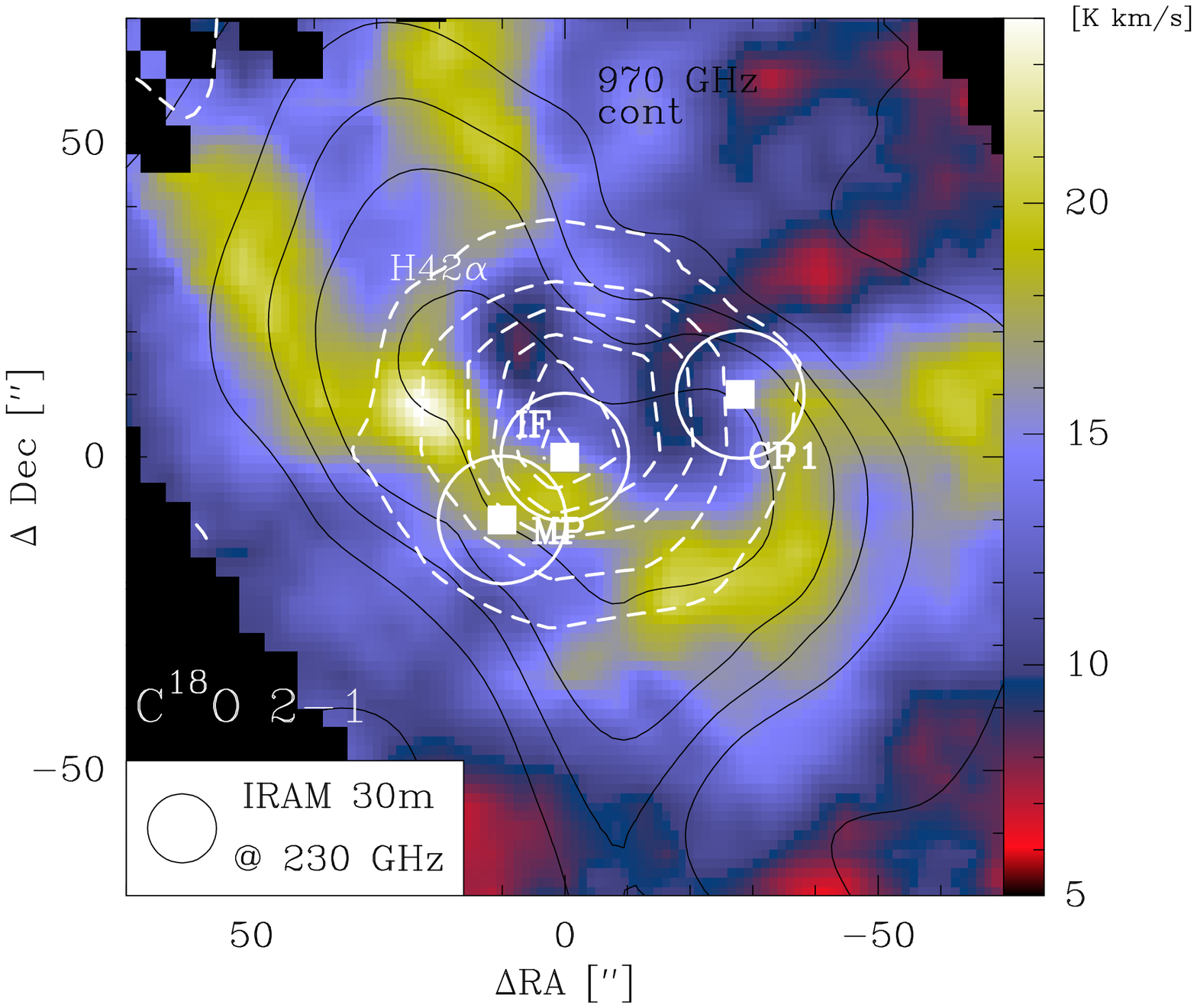}
\caption{In colors, the \CeiO 2-1 integrated intensity in the range [5-15]\kms \citep{pilleri12a}. Squares indicate the positions of the \OHp 1-0 (971\GHz) observations, and the white circles show the HPBW of \herschel at this frequency. 
White dashed contours show the H(42$\alpha$) integrated intensity from 1 to 11\K\kms in linear steps of 2\K\kms.  Black contours represent the relative variation of the continuum emission at 971\GHz from SPIRE-FTS, in 20\% intervals of the peak, 0.6\K.}
\label{fig_obssetup}
\end{figure}

The observations presented here were obtained as part of two  programs of the \herschel Space Observatory \citep{pilbratt10}: the WADI key program \citep[][]{ossenkopf11}, and a follow-up open-time program (PI: P.\,Pilleri). 
We observed the \CHp (1-0, 835\GHz), \OHp (N=1-0, J=2-1,  971\GHz) and \HHOp ($1_{11}-0_{00}$, 1115\GHz) transitions using HIFI to resolve the kinematics of these lines. We also obtained a $2\arcmin\times2\arcmin$ HIFI  on-the-fly (OTF) map of the [\Cp] $^2P_{3/2} \leftarrow P_{1/2}$ line at 1900.5\,GHz. 
The HPBW  of \herschel at these frequencies is 25\arcsec, 22\arcsec, 19\arcsec and 11\arcsec\, for the \CHp, \OHp, \HHOp and [\Cp] lines, respectively. 

In this paper, we present  data taken at three different positions (Fig.\,\ref{fig_obssetup}): the Ionization Front (IF,: RA$_{J2000}$=06h07m46.2s, DEC$_{J2000}=-06^\circ23'08.3''$), the Molecular Peak 1 (MP1) at offset [+10\arcsec, -10\arcsec] relative to the IF, and Continuum Point 1 (CP1) at offset [-28\arcsec, 10\arcsec]. The former two points  have been studied in detail in several molecular and atomic lines \citep[e.g.][]{rizzo03, fuente10, pilleri12a, ginard12, ossenkopf13, pilleri13} and are key references for this source. 
The last point (CP1) was chosen to be the  as far as possible from the 
IF position and still have a significant continuum at the observed frequencies. 

\begin{figure}
\centering
\includegraphics[width=0.7\linewidth]{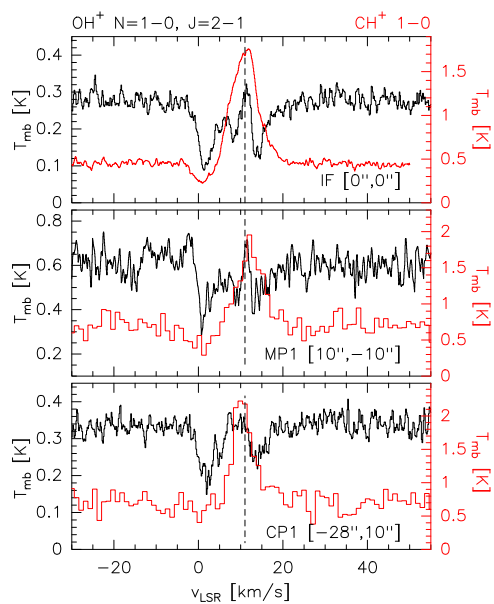}
\caption{Observed spectra of the \CHp (red) and \OHp (black) lines at different position: the Ionization front (IF), the Molecular Peak 1 (MP1) and the Continuum  Point CP1. The continuum levels have been divided by two to correct for the DSB observations. Vertical dashed  lines show the transition velocities, assuming a bulk velocity of the cloud of 11\kms. } 
\label{fig_spectra}
\end{figure}

The \OHp line at all positions and the \HHOp line toward MP1  were observed using single pointing, double beam switch pattern, providing also a reliable measurement of the 971\GHz continuum at all positions and at 1113\GHz toward MP1.  The \CHp observation toward the IF was observed in WADI with a high signal-to-noise (S/N) on a strip centered toward the IF, that also provides a good estimate of the continuum at 835\GHz at this position. 
The remaining  \CHp and \HHOp observations were obtained from a full $2\arcmin\times2\arcmin$ OTF map centered at the IF. Unfortunately,  using OTF on such large areas causes large drifts in the continuum level. To estimate the continuum for these observations, we assume that the relative ratios of the continuum are the same at all positions, and scale the 971\GHz continuum (which is measured at all positions) accordingly.
 We also use the  sparse-sampling continuum map at  971\GHz obtained with SPIRE-FTS within the program SPECHIS-GT (PI: E. Polehampton).

The data were pipelined using  HIPE  \citep{ott10}   9.0 and then Level 2 products were exported to FITS format. Further analysis was performed with the \CLASS/\GILDAS suite \citep{pety05}, and consisted of averaging the two polarizations and conversion to main beam temperature (\Tmb) scale. Finally we divided the observed continuum by 2 since HIFI is a double side band receiver with a sideband ratio of $\sim1$ \citep{roelfsema12}  at these frequencies.

\section{Results}
\label{sec_model}
Figure\,\ref{fig_spectra} displays the detected lines toward the IF, MP1 and CP1.  \HHOp was not detected toward any position (see Table \ref{obsinteg} for the detection limits).  This provides strong limits on the gas molecular fraction \citep{hollenbach12}.
\OHp is detected in absorption at all positions, with two broad features peaking at  blue-shifted  ($\Delta {\rm v} \sim -8\kms$) and red-shifted  ($\Delta {\rm v} \sim 4\kms$) velocities relative to  11\kms, the bulk velocity of the cloud. 
\CHp  is observed in emission in the interval  [5,20]\kms.  Toward IF and MP1, and tentatively toward CP1, it presents an absorption  feature  at blue-shifted velocities ([-5,5]\kms).  

\begin{figure*}
\centering
\includegraphics[width=0.8\linewidth]{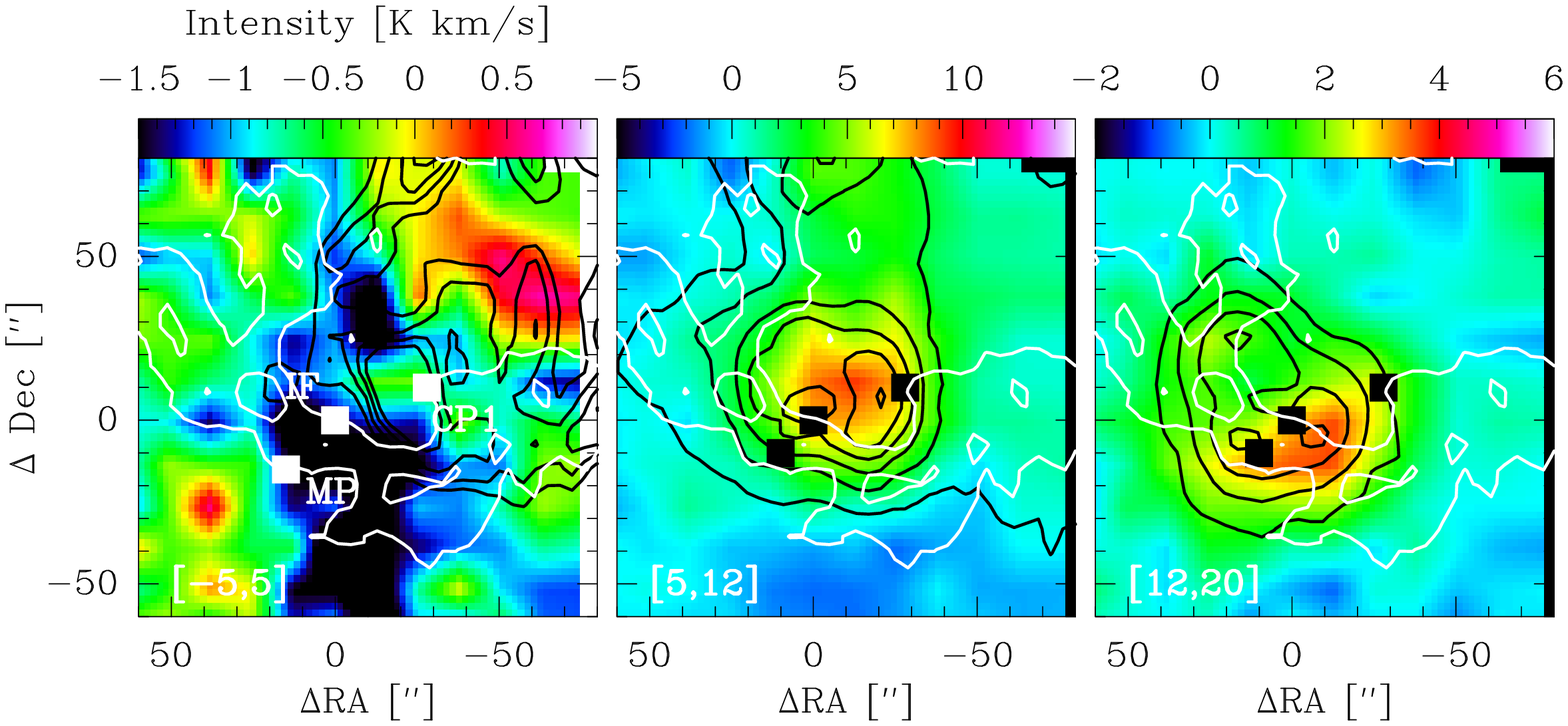}
\caption{Integrated intensity of the $blue$ ([-5, 5]\kms), central ([5,12]\kms) and $red$ ([12, 20]\kms) intervals of the \CHp 1-0 (color scale) and [\Cp] lines (black contours,  starting at 15\Kkms with a step of 5\Kkms for the $blue$ interval, starting at 50\Kkms with a step of 50\Kkms for the central and $red$ intervals).  White contours are the integrated intensity between 5 and 15\kms of the \CeiO 2-1 line.  In the blue interval, the \CHp absorption (dark color) correlates with the molecular cloud.} 
\label{fig_chp_maps}
\end{figure*}

 In Fig. \ref{fig_chp_maps}, we show the integrated intensity maps of the CH$^+$  and [\Cp] lines obtained from our OTF mapping observations in different velocity bins.  The red-shifted emission is clearly related to the \HII region and to the surrounding PDRs, whereas the blue-shifted absorption better correlates with the  \CeiO 2-1 emission, as well as the 971GHz continuum emission (shown in Fig.\,\ref{fig_obssetup}) tracing the bulk of the dust emission associated with the molecular cloud.
 The central velocities [5-12]\kms seem to be associated with the \HII region and the northern, low-UV irradiated PDR described in \citet{ginard12} and \citet{pilleri13}.

 In the following, we focus on the blue- and red-shifted velocities. We define the $blue$ and $red$  intervals as [-5,5]\kms and [12,20]\kms, respectively.  
 In Table \ref{obsinteg} we present a summary of the observational results, detailed below. %

\begin{table*}
\begin{center}
\caption{Observational results.}
\label{obsinteg}
\begin{tabular}{llcccc}
\toprule
&			& &$\int\Tmb d{\rm v}$ &  $\int\tau d{\rm v}$		&  	N(X)			\\
	&	&	&	[\K\kms]	&	[\kms]		&	[$10^{12}$\cmq]\\
\midrule
\CHp		& IF & [-5,5]		&	-0.94		&	2.66		&	8.25\\
\CHp	& MP1	&[-5,5]		&	-2.13		&	4.16		&	12.9\\
\CHp& CP1	&[-5,5]		&	$<$-0.55	&	$<$2.19	&	$<$6.7\\
\midrule
\CHp& IF	&	[12-20]		&	4.26		&	-		&  	74\tablefootmark{(a)}	\\
\CHp& MP1&	[12-20]	&	4.21		&	-		&	72\tablefootmark{(a)}\\
\CHp& CP1&	[12-20]	&	3.05		&	-		&	52\tablefootmark{(a)}		\\
\midrule
\OHp& IF	&	[-5,5]		&	-1.09		&	2.30		&	10.8 \\
\OHp& MP1&	[-5,5]		&	-0.93		&	1.75		&	8.2\\
\OHp& CP1&	[-5,5]		&	-0.64		&	2.40		&	11.3\\
\midrule
\OHp& IF		&[12-20]		&	-0.77		&	1.55		&	7.3\\
\OHp& MP1	&[12-20]&	-0.93		&	1.50		&	7.0\\
\OHp& CP1	&[12-20]	&	-0.39		&	1.36		&	6.4\\
\midrule
\HHOp& IF	&[-5,5], [12-20]		&	-	& 	$<$1.39	&	$<$5.94	\\
\HHOp& MP1	&[-5,5], [12-20]		&	-		&	$<$0.20	&	$<$0.86	 \\
\bottomrule
\end{tabular}
\end{center}
\tablefoot{
 Upper limits are calculated assuming a $3\sigma$ Gaussian profile with a width of 3\kms. \\
\tablefoottext{a} {The} column density of \CHp observed in emission is calculated for $\nhh =\ex{2}{5}$\cmc, $\Tkin = 100\K$, and a line width of 7.5\kms. }
\end{table*}

\paragraph{\CHp}
 Owing to very high critical densities, the \CHp level population  is  out of local thermodynamic equilibrium, thus the column density of the gas associated with the emission line depends on the assumed excitation temperature (\Tex).  As a first approximation, we used the MADEX large velocity gradient (LVG) code \citep{cernicharo12} assuming the typical physical conditions of the PDR \citep[$\nhh \sim \ex{2}{5}$\cmc, $\Tkin \sim 100$\K, ][]{pilleri12a} to estimate the column density of this molecule  at all positions, since the bulk of the \CHp emission is expected to arise in  the most exposed   layers ($A_{\rm V} \lesssim 1$) of the PDR \citep[][]{nagy13}. To fit the observed intensity of the J=1-0  line toward the IF position, the LVG code returns $\Tex \sim10$\K, an opacity of $\tau = 3.2$ and a total column density of \ex{7.4}{13}\cmq for the red-shifted velocity interval.
 On the other extreme, if we consider that the emission stems from  the densest layer \citep[$\nhh = \ex{3}{6}$\cmc, ][]{rizzo03, pilleri12a}, we obtain $\Tex =14\K$ and  $N(\CHp)= \ex{6}{12}\cmq$, corresponding to an opacity of 0.2.

The column density in the $blue$ interval can be calculated with the usual steps for absorption measurements. 
We calculated the opacity of the absorption at each velocity as
$\tau  = -\log ({T_l}/{T_c})$
\noindent where $T_l$ and $T_c$ are  the line and continuum intensities, respectively.  This can be used to estimate the molecular column density  using $N(\CHp) = 3.1\times10^{12} \int \tau {\rm d{\rm v}}$\cmq \citep{godard12}. 
In the expression above, we have assumed that the absorbing \CHp gas is almost entirely in its ground rotational state ($\Tex< 5\K$), which is justified by the relatively large Einstein A$_{ul}$ coefficient for this transition, implying a large critical density ($> 10^6$\cmc).  We also assume that the covering factor over the continuum source is unity. This is reasonable according to the SPIRE map (Fig.\,\ref{fig_obssetup}), although we cannot exclude some clumpiness at smaller scales.

\paragraph{\OHp}
We calculated the integrated opacities $\int \tau d{\rm v}$ of the \OHp line in the different velocity intervals (see Table \ref{obsinteg}).  Since the ground state of \OHp is split into two levels with a negligible energy difference we assume that the population in each level is proportional to their level degeneracies, and included the corresponding correction factor 1.5 in the total column density calculations \citep[see][]{gupta10, neufeld10, gerin10}. This gives a simple expression for the total column density, $N(\OHp) = 4.69\times10^{12} \int \tau d{\rm v}$\cmq. The column density results are reported in Table \ref{obsinteg}.

\paragraph{\HHOp}
 We derived upper limits to the opacity and the column density of \HHOp assuming a  3$\sigma$ detection (calculated at the spectral resolution of 1\kms)  and a line width of 3\kms. Alike \OHp, the ground state of \HHOp is separated into two very close energy levels, and the same correction factor 1.5 must be used to relate the integrated opacity with the column density. 	
This leads to the expression:  
$N(\HHOp) = 4.28\times10^{12} \int \tau  d{\rm v}$ \cmq.  The resulting column densities are reported in  Table \ref{obsinteg}. For CP1, the noise in the spectrum is too high compared to the continuum level to estimate a reliable upper limit to the column density.

\section{Discussion}
\label{sec_disc}

\subsection{Origin of the \OHp and \CHp emission and absorption}

 The column density ratio $N(\OHp)/N(\HHOp)$ toward MP1 is higher than 8 for both the $blue$ and $red$ intervals. We can use this ratio to  estimate an upper limit to the molecular fraction, $f(\HH) = 2\nhh/(n_{\rm H} +2\nhh)$, using the relationship  $[\OHp]/[\HHOp]=0.64+0.12\times(\Tkin/300)^{(-0.5)}/f(\HH)$ \citep{gerin10}.  Assuming \Tkin = 100\K  leads to  $f(\HH)\lesssim 0.03$, {\it i.e.},   the regions producing the absorption consist  of mainly atomic gas, and most likely correspond to the first layer of the PDR or the UV-irradiated walls of an outflow.  Toward the IF and CP1  positions we obtain looser upper limits due to the large $rms$  in the \HHOp spectra.   Because the spatial distribution of most of molecular tracers (such as \CeiO and \thCO) peak toward the molecular bar \citep{pilleri12a} whereas atomic tracers such as [\Cp] have more a circular spatial distribution (Fig. \ref{fig_chp_maps}), we expect that the molecular fraction is lower toward the IF and CP1 compared to the molecular bar, represented by the position MP1. 

The column  density for the $red$ and the $blue$ intervals of \OHp\   is approximately constant toward all three positions.
Even toward CP1, the farthest from the IF and at the  edge of the \HII region,  \OHp  shows spectral profiles and column densities  similar to the IF and CP1 position. This
suggests that  all these absorptions are produced by the same diffuse layers that actually extend  for more than 30\arcsec\, (0.12\,pc)  around the IF.

Several scenarios can be proposed to account for  observed line kinematics. One possible interpretation is that the  \OHp absorption comes from two
 mainly atomic clouds in the line of sight, being both unrelated with the UC\HII. This explanation is unlikely for several reasons. First of all,  the clouds have similar and large linewidths ($\sim$5 \kms) and is unlikely that they originate from two separate clouds. Secondly, the [\Cp] and  CH$^+$ maps (Fig. \ref{fig_chp_maps}) show that this  semi-atomic,  red-shifted gas is spatially constrained to the location of the \HII region.  The similarity of the \CHp and \OHp absorption velocities suggest that \OHp also originates in this region. 
A second possibility is that the absorptions comes from the large scale bipolar outflow associated with this IR cluster \citep{giannakopoulou97, tafalla97,xu06}.
 However, if the absorptions were associated to a bipolar outflow at a certain inclination to the plane of the sky, the $red$ and $blue$ profiles would be significantly different 
toward MP1 and CP1 that are separated by $\sim$43\arcsec, therefore we can discard  this possibility. 
We cannot discard though the presence of an outflow with a perfect symmetrical pole-on configuration, although it is
 quite unlikely to have such a perfectly symmetrical pole-on structure.

 It is more likely, however, that 
 the low molecular fraction  layers are associated with the PDR around the \HII region.
In this case, the  profiles
of the \OHp line can provide important information about
the kinematics of the UCHII region.
 In the picture of an expanding \HII region \citep{jaffe03}, the
values of the velocity centroids of the blue- and red-shifted absorption features would provide an upper limit to the expansion velocity of this layer, $\sim$8 \kms\,  for the front side, and $\sim 4\kms$ for the back side.
The main argument against this interpretation
is that one would expect to see some changes in the separation between the red
and blue absorptions among the observed positions. 
  In the description of \citet{zhu05}, in which the molecular cloud is sustained by the stellar  wind pressure, 
 the atomic gas would be instead entrained in the ionized gas that flows along the walls of the cavity. This latter interpretation  better explains the absence of a difference in the \OHp profiles toward the three positions.  However, an anisotropic expansion or a larger shell could also explain this pattern.
Finally, neither the expansion scenario or the \citet{zhu05} model are able to fully explain the change in the kinematics between the ionized gas  \citep[that peaks at 25\kms, see][]{jaffe03} and the atomic gas (peaking at 15\kms, this paper).  Thus, our data cannot distinguish between these two interpretations.

\begin{figure}
\centering
\includegraphics[width=0.8\linewidth]{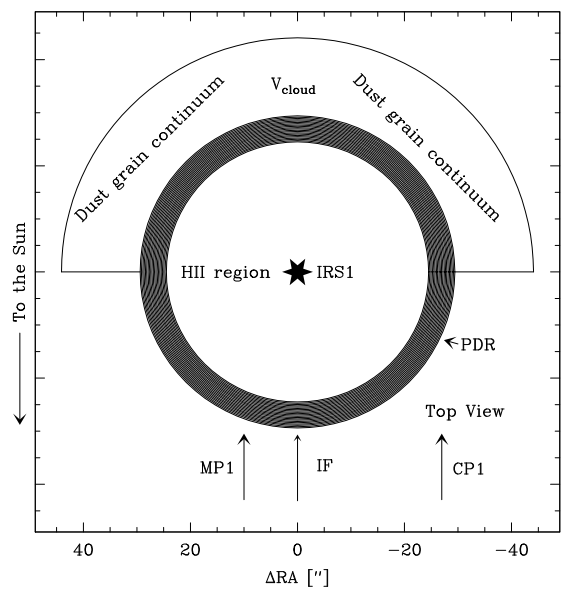}
\caption{Sketch of the Mon~R2 region. The continuum emission is due to the dust grains in the back side of the molecular cloud, and the \OHp and \CHp absorptions arise in the PDR layers.} 
\label{fig_sketch}
\end{figure}

In Fig. \ref{fig_sketch}, we show the sketch of  a model that is consistent with the observations presented here. The  dust grains in the molecular cloud  in the back side (relative to the observer) of the \HII region 
are responsible for the observed continuum emission (see Fig. \ref{fig_obssetup}). 
The \OHp and \CHp\  absorption signatures arise in the  mainly atomic layers of the PDR in front of the molecular
cloud.  
The emission of the high density molecular gas in the back of the \HII region produces the observed absorption-emission profile. 
 The continuum emission must be mainly produced by the dust
behind the \HII region to allow absorption at both blue-shifted
and red-shifted velocities.
This means that the gas density and column densities are larger in the back
than in the front side. 

\subsection{Comparison with chemical models }
\label{sec_meudon}

\begin{figure}
\centering
\includegraphics[angle = 270, width=1\linewidth, trim=1.5cm 0cm 1cm 0cm,clip]{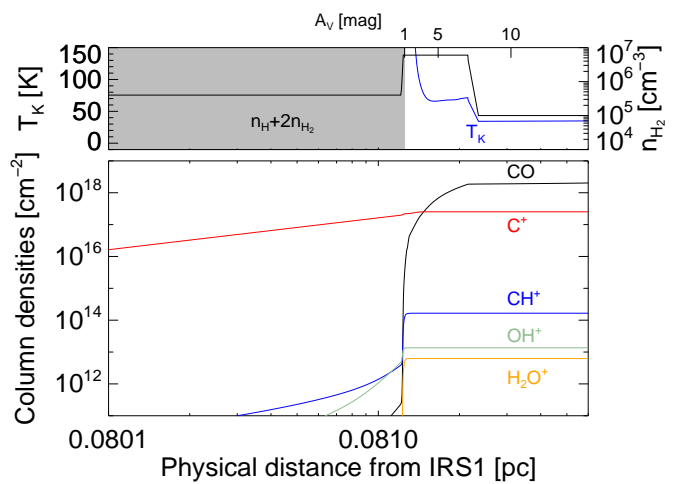}
\caption{Results of the Meudon PDR code, using the same input parameters of \citet{pilleri12a}. The upper panel shows the physical conditions as a function of the distance 
from the IF. The bottom panel shows integrated column densities for the species studied in this work. The shaded area indicates the region where the fraction of molecular to atomic gas, $f(\HH)$, is less than 0.1. } 
\label{fig_meudonres}
\end{figure}

 In our schematic view of Mon~R2, the cloud is constituted by spherically symmetric layers of different densities (Fig.\,\ref{fig_meudonres}). The layers surround the \HII region so that the PDRs are found both in the back and in the 
front of the \HII region. We have used the Meudon PDR code  \citep{lepetit06,goicoechea07,gonzalez08,lebourlot12} version 1.4.3 to compare the column densities derived from the observations to those predicted by gas-phase chemical models.  Since the Meudon code is a 1-D model, these are radial column densities integrated from the beginning of the PDR at $r$=0.08\,pc.  
The input parameters  of the model and the density structure are the same as in \citet{pilleri12a, pilleri13}.

The results of the codes are shown in Fig.\,\ref{fig_meudonres}. In the code, most of the \OHp and \CHp is produced at the transition between the atomic and molecular layers at $A_{\rm V}\sim1$.    The code predicts $N(\OHp) =\ex{1.35}{13}\cmq$ in each side (back and front) of the sphere, in good agreement with the observed values. This  supports that the \OHp absorption is due to the most exposed layers of the PDR. 
Concerning \CHp, the code predicts a column density of \ex{1.6}{13}\cmq\,  in each side of the cloud. This is in  reasonable agreement with the column density obtained from the LVG calculations, considering the large uncertainties in the column density and the simplicity of our sketch model.
 The code predicts also a column density of \HHOp of \ex{6}{12}\cmq, in good agreement with our upper limits toward the IF.

\section{Conclusions}
\label{sec_conclus}

\label{sec_conclus}

We have presented observations of the reactive ions \OHp, \CHp and \HHOp of the Monoceros R2 complex obtained with the HIFI instrument onboard \herschel.  \CHp is observed in emission at the bulk velocity of the region and in absorption at blue-shifted velocities. On the other hand, \OHp is observed in absorption at both blue- and red-shifted velocities, with similar opacities and line profiles at the three observed positions in the cloud. 
 The line profiles of the transitions of \CHp and \OHp observed with HIFI allow us to  constrain the geometry of the region and study the dynamics of the ionized-to-neutral gas conversion layers. 

 Based on our observations, we propose a geometrical model in which the column density of dust grains in the back of the \HII region  (relative to the observer) is higher compared to the front. In this model, the \OHp absorption signatures are associated with the
 layer of mainly atomic gas that surrounds the \HII region. 
Concerning the kinematics, the line profiles of \OHp at all the observed positions are consistent with the scenario of the
atomic
gas flowing along the walls of the cavity with the
ionized gas, as described in \citet{zhu05}.  However, an alternative kinematical scenario consisting in an anisotropic expansion of the UC\HII  cannot be excluded based on our dataset.

\bibliographystyle{aa}
\bibliography{biblio}

\begin{acknowledgements}
 We thank the referee for his useful comments.
The authors thank the spanish MINECO for funding support through the grants
CSD2009-00038 and AYA2009-07304 and AYA2012-32032.
JRG was supported by a Ramon y Cajal contract\\
HIFI has been designed and built by a consortium of institutes and university departments from across
Europe, Canada, and the United States under the leadership of SRON Netherlands Institute for Space
Research, Groningen, The Netherlands and with major contributions from Germany, France, and the US.
Consortium members are: Canada: CSA, U.Waterloo; France: CESR, LAB, LERMA, IRAM; Germany:
KOSMA, MPIfR, MPS; Ireland, NUI Maynooth; Italy: ASI, IFSI-INAF, Osservatorio Astrofisico di Arcetri-
INAF; Netherlands: SRON, TUD; Poland: CAMK, CBK; Spain: Observatorio Astron\'omico Nacional (IGN),
Centro de Astrobiolog\'{\i}a (CSIC-INTA). Sweden: Chalmers University of Technology - MC2, RSS \& GARD;
Onsala Space Observatory; Swedish National Space Board, Stockholm University - Stockholm Observatory;
Switzerland: ETH Zurich, FHNW; USA: Caltech, JPL, NHSC. \\
\end{acknowledgements}

\end{document}